# Using Inclusion Diagrams – as an Alternative to Venn Diagrams – to Determine the Validity of Categorical Syllogisms

Osvaldo Skliar[1]      Ricardo E. Monge[2]      Sherry Gapper[3]

**Abstract:** Inclusion diagrams are introduced as an alternative to using Venn diagrams to determine the validity of categorical syllogisms, and are used here for the analysis of diverse categorical syllogisms. As a preliminary example of a possible generalization of the use of inclusion diagrams, consideration is given also to an argument that includes more than two premises and more than three terms, the classic major, middle and minor terms in categorical syllogisms.

## 1. Introduction

Venn diagrams [1] have been and continue to be used to introduce syllogistics to beginners in the field of logic. These diagrams are perhaps the main technical tool used for didactic purposes to determine the validity of categorical syllogisms.

The objective of this article is to present a different type of diagram, *inclusion diagrams*, to make it possible to reach one of the same objectives for which Venn diagrams were developed. The use of these new diagrams is not only simple and systematic but also very intuitive. It is natural and quite obvious that if one set is included in another set, which in turn, is included in a third set, then the first set is part of the third set. This result, an expression of the transitivity of the inclusion relation existing between sets, is the primary basis of inclusion diagrams.

In technical literature on graphic representations in logic, other authors [2], [3], [4], [5], [6], [7] have pointed toward the usefulness of different types of diagrams for purposes somewhat similar to those addressed here. However, to the best of the authors' knowledge, this graphic method of inclusion diagrams to verify the validity of categorical syllogisms has not been characterized with precision in any of the previous work.

Since it is often culturally enriching in many disciplines, even in scientific fields, to consider problems and their solutions from different points of view, it would appear reasonable and justified to introduce inclusion diagrams, in addition to Venn diagrams and others that already exist.

---

[1] Escuela de Informática, Universidad Nacional, Costa Rica. E-mail: oskliar@costarricense.cr
[2] Escuela de Ciencias de la Computación e Informática. Universidad de Costa Rica, Costa Rica. E-mail: ricardo@mogap.net
[3] Universidad Nacional, Costa Rica. E-mail: sherry.gapper.morrow@una.cr

## 2. Background knowledge

For a clear understanding of this article, it is necessary to be familiar with basic notions of propositional calculus and set theory. Nevertheless, since the main results will also be expressed in words, it is likely that people interested in argumentation who are not familiar with those technical aspects, will still be able to understand most of the ideas presented.

In this article, a) the logical connective of conjunction will be symbolized as $\wedge$; b) the logical connective of implication will be symbolized as $\rightarrow$; and c) the logical connective of equivalence will be symbolized as $\leftrightarrow$.

In logic the symbol $\forall$, known as "universal quantifier", is placed before a variable, such as $x$, to express that "for every" element $x$ of a certain set, the affirmation specified after $\forall x$ is true.

Consider any two sets $C_1$ and $C_2$. The expression $(C_1 \subseteq C_2)$ – that is, the set $C_1$ is included in set $C_2$ – means the following: $\forall x \big((x \in C_1) \rightarrow (x \in C_2)\big)$.

The above statement may be expressed in words: For every element $x$, if that element $x$ belongs to the set $C_1$, then it also belongs to the set $C_2$.

An expression equivalent to $(C_1 \subseteq C_2)$ is: $C_1$ is a subset of $C_2$.

Given the characterization provided of the notion of the inclusion of one set within another set, it is essential to clarify that every set is included in itself; that is, every set is a subset of itself. In other words, the statement $(C_1 \subseteq C_2)$ does not exclude the possibility that $C_1$ and $C_2$ are the same set: $C_1 = C_2$.

Recall that the "universal set" (or the "universe of discourse"), usually referred to by using $\bigcup$, is the set to which all the elements considered belong in a given context.

The complementary set of any set $C_1$ will be symbolized as $\overline{C_1}$. Take into account that all the elements of $\bigcup$ which do not belong $C_1$ belong to $\overline{C_1}$.

To symbolize a subset included in $C_1$, the symbol $SC_1$ will be used; of course, $(SC_1 \subseteq C_1)$.

Four results from set theory (I, II, III, IV) which will be used in this article will be specified below.

I. $\overline{\overline{C_1}} = C_1$ (1)

In words: The complement of the complement of a set is equal to that set.

II. $(C_1 \subseteq C_2) \leftrightarrow (\overline{C_2} \subseteq \overline{C_1})$ (2)

In words: The propositions $(C_1 \subseteq C_2)$ and $(\overline{C_2} \subseteq \overline{C_1})$ are equivalents. This means that these propositions have the same truth value. If one of these propositions is true, the other one also is true; and if one of these propositions is false, then the other also is false.

III. $(SC_1 \subseteq C_2) \leftrightarrow (SC_2 \subseteq C_1)$; $SC_1 = SC_2$ (3)

In words: If a subset $SC_1$ of the set $C_1$ is included in a set $C_2$ (that is, if $SC_1$ is also a subset of $C_2$), then the subset $SC_2$ of $C_2$ is included in $C_1$. It suffices to consider the subset $SC_2$ of $C_2$, such that $SC_2 = SC_1$, to note that $SC_2$ is also a subset of $C_1$. In addition, if a subset $SC_2$ of the set $C_2$ is included in a subset $C_1$ (that is, if $SC_2$ is also a subset of $C_1$), then a specific subset $SC_1$ of $C_1$ is included in $C_2$. It suffices to consider the subset $SC_1$ of $C_1$, such that $SC_1 = SC_2$, to see that $SC_1$ is also a subset of $C_2$. Thus the propositions $(SC_1 \subseteq C_2)$ and $(SC_2 \subseteq C_1)$ are equivalent, where $SC_1 = SC_2$.

IV. $\quad ((C_1 \subseteq C_2) \wedge (C_2 \subseteq C_3)) \rightarrow (C_1 \subseteq C_3)$ $\qquad\qquad$ (4)

In words: If the propositions $(C_1 \subseteq C_2)$ and $(C_2 \subseteq C_3)$ are true, then the proposition $(C_1 \subseteq C_3)$ is true. The validity of (4), which expresses the transitivity of the inclusion relation between sets, is very intuitive. If any element of $C_1$ also belongs to $C_2$, and any element of $C_2$ also belongs to $C_3$, then it is clear that any element of $C_1$ also belongs to $C_3$.

It must be emphasized that the accuracy of the preceding statements (1), (2), (3), and (4) is certainly independent of the denominations $C_1$, $C_2$, and $C_3$, chosen for the sets considered. Suppose that those denominations were changed to $C_{27}$, $C_6$, and $C_{19}$, respectively. Then those statements would take on the following forms:

$$\overline{\overline{C_{27}}} = C_{27} \qquad\qquad (1)$$

$$(C_{27} \subseteq C_6) \leftrightarrow (\overline{C_6} \subseteq \overline{C_{27}}) \qquad\qquad (2)$$

$$(SC_{27} \subseteq C_6) \leftrightarrow (SC_6 \subseteq C_{27}) \, ; \; SC_{27} = SC_6 \qquad\qquad (3)$$

$$((C_{27} \subseteq C_6) \wedge (C_6 \subseteq C_{19})) \rightarrow (C_{27} \subseteq C_{19}) \qquad\qquad (4)$$

One characteristic of all the sets discussed in the sections below is that the elements belonging to each set have a certain property in common, such as that of being men, being wise, being stones or being rivers.

## 3. Brief review of notions of categorical propositions and categorical syllogisms

### 3.1 Categorical propositions

Categorical propositions can be considered assertions (about sets) which affirm or negate that one set is partially or totally included in another.

Four examples of categorical propositions are given below. Each corresponds to one of the four characteristic forms of this type of proposition.

1. All generals are brave.
2. No general is brave.
3. Some generals are brave.
4. Some generals are not brave.

The above statements refer to two sets: $C_1$, the set of all generals; and $C_2$, the set of all brave persons. These sets can be referred to in the following abbreviated form:

$C_1$: generals; $C_2$: brave persons

Given the notation introduced, the following four examples of categorical propositions are expressed as follows:

1. All $C_1$'s are $C_2$'s.

In words: Every element belonging to $C_1$ also belongs to $C_2$, or in the terminology of set theory: $C_1 \subseteq C_2$.

The first proposition is an example of those known as universal affirmative propositions.

2. No $C_1$ is a $C_2$.

In words: No element of $C_1$ also belongs to $C_2$. Any element of $\bigcup$ that does not belong to $C_2$ must belong to $\overline{C_2}$ (the complement of $C_2$). Therefore, using set theory symbols, proposition 2 is expressed as: $C_1 \subseteq \overline{C_2}$.

This second proposition is an example of those known as universal negative propositions.

3. Some $C_1$'s are $C_2$'s.

In words: Some elements of $C_1$ also belong to $C_2$. If the notation introduced in section 2 is used, proposition 3, expressed using set theory symbols, is $SC_1 \subseteq C_2$.

This third proposition is an example of those known as particular affirmative propositions.

4. Some $C_1$'s are not $C_2$'s.

In words: Some elements of $C_1$ do not belong to $C_2$. As indicated above, any element of $\bigcup$ that does not belong to $C_2$ must belong to $\overline{C_2}$. Thus using set theory symbols and the notation introduced in section 2, proposition 4 is expressed as: $SC_1 \subseteq \overline{C_2}$.

It is clear that instead of considering that $C_1$ and $C_2$ are respectively sets of generals and brave persons, any other sets may be considered, such as the set of tigers and that of fierce animals.

## 3.2 Typical categorical syllogisms

A syllogism is a type of deductive reasoning in which a proposition (known as a *conclusion*) is inferred, or deduced, from two propositions known as *premises*.

Consider, for example, the following syllogism:

    1) All engineers are pragmatic.
    2) Some engineers are wealthy.

∴  3) Some wealthy persons are pragmatic.

It can be observed that this categorical syllogism is composed of three categorical propositions which have been numbered 1), 2), and 3). The first two propositions are the premises of the syllogism and the third proposition is the conclusion of that syllogism.

The symbol $\therefore$ (preceding the third proposition) means "therefore".

The predicate term of the conclusion (*pragmatic*) is known as the *major term* of the syllogism and the subject term of the conclusion, wealthy persons, is called the *minor term* of the syllogism. The major premise of the syllogism is that which contains the major term. In this case the first premise is the major premise since it contains the major term (*pragmatic*) and the second premise is the minor premise, since it contains the minor term (*wealthy persons*).

The third term of the syllogism does not appear in the conclusion but it does appear in each of the premises. This third term is called the *middle term*. In the syllogism considered this third term is *engineers*.

Each term of a syllogism may be made to correspond to a particular set. Thus, in this syllogism the major term (pragmatic) corresponds to the set $C_3$ of all pragmatic persons, and may be abbreviated as follows:

$C_3$: pragmatic persons

The minor term of the syllogism (wealthy persons) corresponds to the set $C_1$ of all wealthy persons. When abbreviated, $C_1$ is characterized as:

$C_1$: wealthy persons

The middle term of the syllogism considered (engineers) can be made to correspond to set $C_2$, that of all the engineers. This can be abbreviated as:

$C_2$: engineers

If attention is given to what sets $C_1$, $C_2$, and $C_3$ are, the respective syllogism may be expressed as:

$$\text{1)} \quad C_2 \subseteq C_3$$
$$\text{2)} \quad SC_2 \subseteq C_1$$
$$\therefore \quad \text{3)} \quad SC_1 \subseteq C_3$$

The validity of this syllogism will be tested in section 4.

A categorical syllogism is deemed to be presented in a standard way if the first of the three propositions of which it is composed is the major premise, the second of them is the minor premise, and the third the conclusion.

## 4. Inclusion diagrams and tests of the validity of diverse categorical syllogisms

Inclusion diagrams will be introduced in this section. These diagrams make it possible to discern, clearly and intuitively, the validity of categorical syllogisms that are in fact valid. If a syllogism is not valid, it is impossible to construct the respective inclusion diagram.

A number of examples of valid categorical syllogisms expressed in a standard form will be addressed below. The corresponding inclusion diagram will be provided for each.

Example 1

       1) All engineers are pragmatic.
       2) Some engineers are wealthy.
∴    3) Some wealthy persons are pragmatic.

Consider the following sets:

$C_1$: wealthy persons;         $C_2$: engineers;     $C_3$: pragmatic persons
$SC_1$: some wealthy persons;    $SC_2$: some engineers

Given the notation introduced, the syllogism considered may be expressed as:

    1)    $C_2 \subseteq C_3$
    2)   $SC_2 \subseteq C_1$
∴   3)   $SC_1 \subseteq C_3$

Note that, due to (3) in section 2, there is a proposition equivalent to the second premise:

$$(SC_2 \subseteq C_1) \leftrightarrow (SC_1 \subseteq C_2)\,;\ SC_2 = SC_1$$

In words: If the second premise is true, as it is here, then $(SC_1 \subseteq C_2)$ also is true. If the second premise were false, then $(SC_1 \subseteq C_2)$ also would be false. Therefore, the syllogism considered can be reformulated as:

    1)    $C_2 \subseteq C_3$
    2)   $SC_1 \subseteq C_2$
∴   3)   $SC_1 \subseteq C_3$

The conjunction of the two premises is the following proposition:

$$(SC_1 \subseteq C_2) \wedge (C_2 \subseteq C_3)$$

Due to (4), in section 2, the following proposition is valid:

$$\big((SC_1 \subseteq C_2) \wedge (C_2 \subseteq C_3)\big) \rightarrow \big(SC_1 \subseteq C_3\big)$$

In words: If the propositions $(SC_1 \subseteq C_2)$ and $(C_2 \subseteq C_3)$ are true (i.e., the second and first premises of this syllogism, respectively), then the proposition $(SC_1 \subseteq C_3)$ also is true. That is, given the meanings of $SC_1$ and $C_3$, the proposition "Some wealthy persons are pragmatic" is true. This proposition is the conclusion of the syllogism considered.

The above reasoning may be shown graphically as follows:

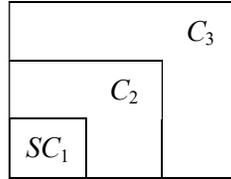

Figure 1. Inclusion diagram for the categorical syllogism in example 1.
See explanation below.

In figure 1, it can be clearly seen that given $(SC_1 \subseteq C_2)$ (a proposition equivalent to the minor premise) and the major premise $(C_2 \subseteq C_3)$, the proposition $(SC_1 \subseteq C_3)$ – the conclusion of the syllogism – is true.

It is irrelevant that the sets $SC_1$, $C_2$, and $C_3$ happen to have been represented by using three *rectangles* such that the area corresponding to $SC_1$ is included within the area corresponding to $C_2$, and the area corresponding to $C_2$ is included within the area of $C_3$. These three sets could have been represented by areas in circles, triangles or other geometrical figures. What is relevant here is the existence of relations of inclusion, as indicated.

Essentially, the boundary of $SC_1$ could coincide with that of $C_2$, in which case the following equality would be valid: $SC_1 = C_2$. The boundary of $C_2$ could also coincide with that of $C_3$, in which case the following equality would be valid: $C_2 = C_3$. The syllogism considered in example 1 remains valid regardless of whether both $SC_1 = C_2$ and $C_2 = C_3$ are valid, or only one of them is valid. A similar consideration is applicable to the rest of the examples of categorical syllogisms to be discussed.

In the case of categorical syllogisms, only three different areas appear in the corresponding inclusion diagrams. In section 6, where this type of diagram is applied to an argument different from that of a categorical syllogism, more than three areas may appear.

Example 2

      1) No intellectual is superstitious.
      2) Some French persons are intellectuals.

∴   3) Some French persons are not superstitious.

Consider the following sets:

$C_1$: French persons;     $C_2$: intellectuals;     $C_3$: superstitious persons
$SC_1$: some French persons

Given the notation introduced, the syllogism considered may be expressed as:

$\quad$ 1) $\quad C_2 \subseteq \overline{C_3}$

$\quad$ 2) $\quad SC_1 \subseteq C_2$

$\therefore \quad$ 3) $\quad SC_1 \subseteq \overline{C_3}$

The conjunction of $(SC_1 \subseteq C_2)$, the minor premise, and $(C_2 \subseteq \overline{C_3})$, the major premise, is the following proposition:

$$(SC_1 \subseteq C_2) \wedge (C_2 \subseteq \overline{C_3})$$

Due to (4), the following proposition is valid:

$$\left((SC_1 \subseteq C_2) \wedge (C_2 \subseteq \overline{C_3})\right) \rightarrow \left(SC_1 \subseteq \overline{C_3}\right)$$

In words: If the propositions $(SC_1 \subseteq C_2)$ and $(C_2 \subseteq \overline{C_3})$ are true, then the proposition $(SC_1 \subseteq \overline{C_3})$ also is true. That is, given the meanings of $SC_1$ and $C_3$, the proposition "Some French persons are not superstitious" is true. This proposition is the conclusion of the syllogism considered.

The above reasoning may be shown graphically as follows:

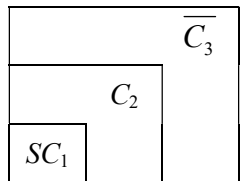

Figure 2. Inclusion diagram for the categorical syllogism in example 2.

Example 3

$\quad$ 1) All men are rational.

$\quad$ 2) All Spaniards are men.

$\therefore \quad$ 3) All Spaniards are rational.

Consider the following sets:

$C_1$: Spaniards;     $C_2$: men;     $C_3$: rational men

Given the notation introduced, the syllogism considered may be expressed as:

$$\begin{array}{ll} 1) & C_2 \subseteq C_3 \\ 2) & C_1 \subseteq C_2 \\ \therefore \quad 3) & C_1 \subseteq C_3 \end{array}$$

The conjunction of $(C_1 \subseteq C_2)$, the minor premise, and $(C_2 \subseteq C_3)$, the major premise, is the following proposition: $(C_1 \subseteq C_2) \wedge (C_2 \subseteq C_3)$.

Due to (4), the following proposition is valid:

$$\big((C_1 \subseteq C_2) \wedge (C_2 \subseteq C_3)\big) \rightarrow \big(C_1 \subseteq C_3\big)$$

In words: If the propositions $(C_1 \subseteq C_2)$, the minor premise, and $(C_2 \subseteq C_3)$, the major premise, are true, then the proposition $(C_1 \subseteq C_3)$ also is true. That is, given the meanings of $C_1$ and $C_3$, the proposition "All Spaniards are rational" is true. This proposition is the conclusion of the syllogism considered.

The above reasoning may be shown graphically as follows:

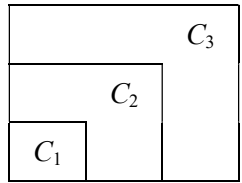

Figure 3. Inclusion diagram for the categorical syllogism in example 3.

Example 4

    1) All mathematicians are clever.
    2) Some Germans are mathematicians.
$\therefore$  3) Some Germans are clever.

Consider the following sets:

$C_1$: Germans;       $C_2$: mathematicians;      $C_3$: clever persons
$SC_1$: some Germans

Given the notation introduced, the syllogism considered may be expressed as:

$$\begin{array}{ll} 1) & C_2 \subseteq C_3 \\ 2) & SC_1 \subseteq C_2 \\ \therefore \quad 3) & SC_1 \subseteq C_3 \end{array}$$

The conjunction of $(SC_1 \subseteq C_2)$, the minor premise, and $(C_2 \subseteq C_3)$, the major premise, is the following proposition:

$(SC_1 \subseteq C_2) \land (C_2 \subseteq C_3)$

Due to (4), the following proposition is valid:

$$\big((SC_1 \subseteq C_2) \land (C_2 \subseteq C_3)\big) \to \big(SC_1 \subseteq C_3\big)$$

If the propositions $(SC_1 \subseteq C_2)$ and $(C_2 \subseteq C_3)$ are true, then the proposition $(SC_1 \subseteq C_3)$ also is true. That is, given the meanings of $SC_1$ and $C_3$, the proposition "Some Germans are clever" is true. This proposition is the conclusion of the syllogism considered.

The above reasoning may be shown graphically as follows:

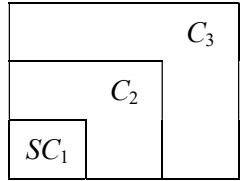

Figure 4. Inclusion diagram for the categorical syllogism in example 4.

Example 5

 1) All mammals are vertebrates.
 2) Some animals are not vertebrates.
∴ 3) Some animals are not mammals.

Consider the following sets:

 $C_1$: animals;   $C_2$: vertebrates;   $C_3$: mammals
$SC_1$: some animals

Given the notation introduced, the syllogism considered may be expressed as:

 1) $C_3 \subseteq C_2$
 2) $SC_1 \subseteq \overline{C_2}$
∴ 3) $SC_1 \subseteq \overline{C_3}$

Due to (2), in section 2, the following proposition is valid:

$$(C_3 \subseteq C_2) \leftrightarrow (\overline{C_2} \subseteq \overline{C_3})$$

Given that the major premise $(C_3 \subseteq C_2)$ is equivalent to the proposition $(\overline{C_2} \subseteq \overline{C_3})$, the syllogism considered can be reformulated as:

1)  $(\overline{C_2} \subseteq \overline{C_3})$

2)  $SC_1 \subseteq \overline{C_2}$

$\therefore$  3)  $SC_1 \subseteq \overline{C_3}$

The conjunction of the propositions if $(SC_1 \subseteq \overline{C_2})$ and $(\overline{C_2} \subseteq \overline{C_3})$ is the following proposition:

$$(SC_1 \subseteq \overline{C_2}) \wedge (\overline{C_2} \subseteq \overline{C_3})$$

Due to (4), the following proposition is valid:

$$\left((SC_1 \subseteq \overline{C_2}) \wedge (\overline{C_2} \subseteq \overline{C_3})\right) \rightarrow \left(SC_1 \subseteq \overline{C_3}\right)$$

In words: If both premises of the syllogism considered, once reformulated, are true, then the proposition $(SC_1 \subseteq \overline{C_3})$ also is true. That is, given the meanings of $SC_1$ and $C_3$, the proposition "Some animals are not mammals" is true. This proposition is the conclusion of the syllogism considered.

The above reasoning may be shown graphically as follows:

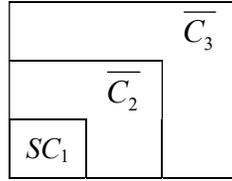

Figure 5. Inclusion diagram for the categorical syllogism in example 5.

Example 6

    1) Some Argentineans are not agnostics.

    2) All Argentineans are South Americans.

$\therefore$  3) Some South Americans are not agnostics.

Consider the following sets:

$C_1$: South Americans;        $C_2$: Argentineans;        $C_3$: agnostics

$SC_1$: some South Americans;   $SC_2$: some Argentineans

Given the notation introduced, the syllogism considered may be expressed as:

1)  $SC_2 \subseteq \overline{C_3}$

2)  $C_2 \subseteq C_1$

$$\therefore \quad 3) \quad SC_1 \subseteq \overline{C_3}$$

Due to (3), in section 2, the first premise is equivalent to the proposition $(S\overline{C_3} \subseteq C_2)$:

$$(SC_2 \subseteq \overline{C_3}) \leftrightarrow (S\overline{C_3} \subseteq C_2)\,;\ SC_2 = S\overline{C_3}$$

The assertion that some Argentineans are not agnostics (or rather the assertion that those Argentineans are not agnostics) is equivalent to the assertion that some non-agnostics are Argentineans. The syllogism considered can be reformulated then as follows:

$$1) \quad S\overline{C_3} \subseteq C_2$$
$$2) \quad C_2 \subseteq C_1$$
$$\therefore \quad 3) \quad SC_1 \subseteq \overline{C_3}$$

The conjunction of the two premises of the syllogism considered, once reformulated, is the following proposition:

$$(S\overline{C_3} \subseteq C_2) \wedge (C_2 \subseteq C_1)$$

Due to (4), the following proposition is valid:

$$\left((S\overline{C_3} \subseteq C_2) \wedge (C_2 \subseteq C_1)\right) \rightarrow \left(S\overline{C_3} \subseteq C_1\right)$$

In addition, due to (3), this proposition is valid:

$$(S\overline{C_3} \subseteq C_1) \leftrightarrow (SC_1 \subseteq \overline{C_3})\,;\ S\overline{C_3} = SC_1$$

In words: If both premises of this reformulated syllogism are true, then the proposition $(S\overline{C_3} \subseteq C_1)$ also is true. That is, given the meanings of $C_3$ and $C_1$, the proposition "Some non-agnostics are South Americans" is true. Due to (3), this proposition is equivalent to "Some South Americans are non-agnostics". That is, "Some South Americans are not agnostics"; and this proposition is the conclusion of the syllogism considered.

The above reasoning may be shown graphically as follows:

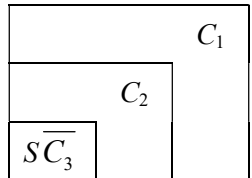

Figure 6. Inclusion diagram for the categorical syllogism in example 6.

Recall the equivalence $(S\overline{C_3} \subseteq C_1) \leftrightarrow (SC_1 \subseteq \overline{C_3})$.

Example 7

     1) All sculptors are artists.
     2) No artist is a fossil.
$\therefore$   3) No fossil is a sculptor.

Consider the following sets:

$C_1$: fossils;     $C_2$: artists;     $C_3$: sculptors

Given the notation introduced, the syllogism considered may be expressed as:

    1)   $C_3 \subseteq C_2$
    2)   $C_2 \subseteq \overline{C_1}$
$\therefore$  3)   $C_1 \subseteq \overline{C_3}$

The conjunction of the two premises is the following proposition:

$$(C_3 \subseteq C_2) \wedge (C_2 \subseteq \overline{C_1})$$

Due to (4), the following proposition is valid:

$$\left((C_3 \subseteq C_2) \wedge (C_2 \subseteq \overline{C_1})\right) \rightarrow \left(C_3 \subseteq \overline{C_1}\right)$$

In words: If both premises of this reformulated syllogism are true, then the proposition $(C_3 \subseteq \overline{C_1})$ also is true. Due to (2) and (1), in section 2, this proposition is equivalent to the proposition $(C_1 \subseteq \overline{C_3})$, which is the conclusion of the syllogism considered.

The above reasoning may be shown graphically as follows:

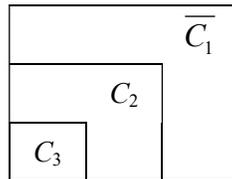

Figure 7. Inclusion diagram for the categorical syllogism in example 7.
Recall the following equivalence: $(C_3 \subseteq \overline{C_1}) \leftrightarrow (C_1 \subseteq \overline{C_3})$

Example 8

1) All multimillionaires are magnanimous.
2) No fanatic is magnanimous.

∴   3) No fanatic is a multimillionaire.

Consider the following sets:

$C_1$: fanatics;        $C_2$: magnanimous persons;        $C_3$: multimillionaires.

Given the notation introduced, the syllogism considered may be expressed as:

1)   $C_3 \subseteq C_2$

2)   $C_1 \subseteq \overline{C_2}$

∴   3)   $C_1 \subseteq \overline{C_3}$

Given (2) and (1), the second premise is equivalent to the following proposition: $(C_2 \subseteq \overline{C_1})$. Therefore, the above syllogism may be reformulated as:

1)   $C_3 \subseteq C_2$

2)   $C_2 \subseteq \overline{C_1}$

∴   3)   $C_1 \subseteq \overline{C_3}$

The conjunction of the two premises of the syllogism considered, once reformulated, is the following:

$$(C_3 \subseteq C_2) \wedge (C_2 \subseteq \overline{C_1})$$

Due to (4), the following proposition is valid:

$$\left( (C_3 \subseteq C_2) \wedge (C_2 \subseteq \overline{C_1}) \right) \to \left( C_3 \subseteq \overline{C_1} \right)$$

In words: If both premises are true, then the proposition $(C_3 \subseteq \overline{C_1})$ also is true. Due to (2) and (1), this proposition is equivalent to the proposition $(C_1 \subseteq \overline{C_3})$; or in words: "No fanatic is a multimillionaire". This is the conclusion of the syllogism considered.

The above reasoning may be shown graphically as follows:

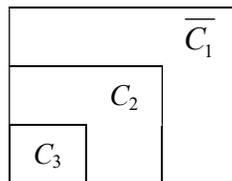

Figure 8. Inclusion diagram for the categorical syllogism in example 8.

Recall the following equivalence: $(C_3 \subseteq \overline{C_1}) \leftrightarrow (C_1 \subseteq \overline{C_3})$

Example 9

    1) No humanist is corrupt.
    2) All despots are corrupt.
$\therefore$   3) No despot is a humanist.

Consider the following sets:

$C_1$: despots;       $C_2$: corrupt persons;       $C_3$: humanists

Given the notation introduced, the syllogism considered may be expressed as:

    1)  $C_3 \subseteq \overline{C_2}$
    2)  $C_1 \subseteq C_2$
$\therefore$  3)  $C_1 \subseteq \overline{C_3}$

Due to (2) and (1), the following equivalence is valid:

$(C_3 \subseteq \overline{C_2}) \leftrightarrow (C_2 \subseteq \overline{C_3})$

Therefore, the syllogism can be reformulated as:

    1)  $C_2 \subseteq \overline{C_3}$
    2)  $C_1 \subseteq C_2$
$\therefore$  3)  $C_1 \subseteq \overline{C_3}$

The conjunction of the two premises of the above syllogism, once reformulated, is the following proposition:

$(C_1 \subseteq C_2) \wedge (C_2 \subseteq \overline{C_3})$

Due to (4), the following proposition is valid:

$\left((C_1 \subseteq C_2) \wedge (C_2 \subseteq \overline{C_3})\right) \rightarrow \left(C_1 \subseteq \overline{C_3}\right)$

If both premises are true, then the proposition $(C_1 \subseteq \overline{C_3})$ also is true. This proposition is the conclusion of the syllogism considered.

The above reasoning may be shown graphically as follows:

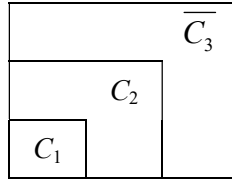

Figure 9. Inclusion diagram for the categorical syllogism in example 9.

Example 10

     1) All logicians are wise.
     2) Some logicians are Poles.

$\therefore$   3) Some Poles are wise.

Consider the following sets:

  $C_1$: Poles;         $C_2$: logicians;     $C_3$: wise persons
$SC_1$: some Poles;    $SC_2$: some logicians

Given the notation introduced, the syllogism considered may be expressed as:

    1)   $C_2 \subseteq C_3$
    2)  $SC_2 \subseteq C_1$

$\therefore$   3)  $SC_1 \subseteq C_3$

Due to (3), the following equivalence is valid:

$(SC_2 \subseteq C_1) \leftrightarrow (SC_1 \subseteq C_2)\,;\ SC_1 = SC_2$

Therefore, the syllogism can be reformulated as:

    1)   $C_2 \subseteq C_3$
    2)  $SC_1 \subseteq C_2$

$\therefore$   3)  $SC_1 \subseteq C_3$

The conjunction of the two premises is the following proposition:

$(SC_1 \subseteq C_2) \wedge (C_2 \subseteq C_3)$

Due to (4), the following proposition is valid:

$\big((SC_1 \subseteq C_2) \wedge (C_2 \subseteq C_3)\big) \rightarrow \big(SC_1 \subseteq C_3\big)$

If both premises are true, then the proposition $(SC_1 \subseteq C_3)$ also is true. This proposition is the conclusion of the syllogism considered.

The above reasoning may be shown graphically as follows:

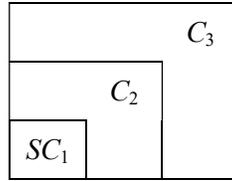

Figure 10. Inclusion diagram for the categorical syllogism in example 10.

Example 11

     1) All sportsmen are long-lived.
     2) Some sportsmen are Ukrainians.

∴   3) Some Ukrainians are long-lived.

Consider the following sets:

  $C_1$: Ukrainians;        $C_2$: sportsmen;        $C_3$: long-lived persons
$SC_1$: some Ukrainians;    $SC_2$: some sportsmen

Given the notation introduced, the syllogism considered may be expressed as:

     1)   $C_2 \subseteq C_3$

     2)  $SC_2 \subseteq C_1$

∴   3)  $SC_1 \subseteq C_3$

Due to (3), the following proposition is valid:

$(SC_2 \subseteq C_1) \leftrightarrow (SC_1 \subseteq C_2)\,;\; SC_1 = SC_2$

Therefore the syllogism can be reformulated as:

     1)   $C_2 \subseteq C_3$

     2)  $SC_1 \subseteq C_2$

∴   3)  $SC_1 \subseteq C_3$

The conjunction of the two premises of the reformulated syllogism is the following proposition:

$(SC_1 \subseteq C_2) \wedge (C_2 \subseteq C_3)$

Due to (4), the following proposition is valid:

$$\big((SC_1 \subseteq C_2) \wedge (C_2 \subseteq C_3)\big) \rightarrow \big(SC_1 \subseteq C_3\big)$$

If both premises are true, then the proposition $(SC_1 \subseteq C_3)$ also is true. This proposition is the conclusion of the syllogism considered.

The above reasoning may be shown graphically as follows:

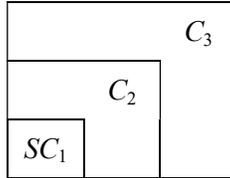

Figure 11. Inclusion diagram for the categorical syllogism in example 11.

Example 12

    1) Some mammals are dogs.
    2) All mammals are vertebrates.
∴   3) Some vertebrates are dogs.

Consider the following sets:

 $C_1$: vertebrates;        $C_2$: mammals;    $C_3$: dogs
$SC_1$: some vertebrates;    $SC_2$: some mammals

Given the notation introduced, the syllogism considered may be expressed as:

    1)   $SC_2 \subseteq C_3$

    2)   $C_2 \subseteq C_1$

∴   3)   $SC_1 \subseteq C_3$

Due to (3), the following equivalence is valid:

$$(SC_2 \subseteq C_3) \leftrightarrow (SC_3 \subseteq C_2)\,;\ SC_2 = SC_3$$

Therefore, the syllogism can be reformulated as:

    1)   $SC_3 \subseteq C_2$

    2)   $C_2 \subseteq C_1$

∴   3)   $SC_1 \subseteq C_3$

The conjunction of the two premises of the syllogism, once reformulated, is the following proposition:

$$(SC_3 \subseteq C_2) \wedge (C_2 \subseteq C_1)$$

Due to (4), the following proposition is valid:

$$\big((SC_3 \subseteq C_2) \wedge (C_2 \subseteq C_1)\big) \rightarrow \big(SC_3 \subseteq C_1\big)$$

If both premises are true, then the proposition $(SC_3 \subseteq C_1)$ also is true. This proposition, due to (3) is equivalent to the proposition $(SC_1 \subseteq C_3)$. This proposition is the conclusion of the syllogism considered.

The above reasoning may be shown graphically as follows:

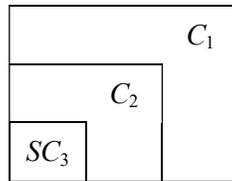

Figure 12. Inclusion diagram for the categorical syllogism in example 12. Recall the following equivalence: $(SC_3 \subseteq C_1) \leftrightarrow (SC_1 \subseteq C_3)$.

Example 13

    1) Some entrepreneurs are public accountants.
    2) All public accountants are sensible.
∴  3) Some sensible persons are entrepreneurs.

Consider the following sets:

$C_1$: sensible persons;        $C_2$: public accountants;        $C_3$: entrepreneurs
$SC_1$: some sensible persons;                     $SC_3$: some entrepreneurs

Given the notation introduced, the syllogism considered may be expressed as:

    1)   $SC_3 \subseteq C_2$

    2)   $C_2 \subseteq C_1$

∴  3)   $SC_1 \subseteq C_3$

The conjunction of the two premises is the following proposition:

$$(SC_3 \subseteq C_2) \wedge (C_2 \subseteq C_1)$$

Due to (4), the following proposition is valid:

$$\big((SC_3 \subseteq C_2) \wedge (C_2 \subseteq C_1)\big) \to \big(SC_3 \subseteq C_1\big)$$

If both premises are true, then the proposition $(SC_3 \subseteq C_1)$ also is true. This proposition, due to (3), is equivalent to the proposition $(SC_1 \subseteq C_3)$, which is the conclusion of the syllogism considered.

The above reasoning may be shown graphically as follows:

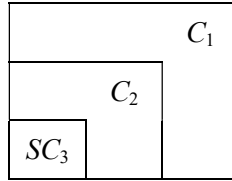

Figure 13. Inclusion diagram for the categorical syllogism in example 13. Recall the following equivalence: $(SC_3 \subseteq C_1) \leftrightarrow (SC_1 \subseteq C_3)$.

Example 14

     1) No artist is Neo-Kantian,
     2) Some Germans are Neo-Kantians.

∴   3) Some Germans are not artists.

Consider the following sets:

 $C_1$: Germans;       $C_2$: Neo-Kantians;     $C_3$: artists;
$SC_1$: some Germans.

Given the notation introduced, the syllogism considered may be expressed as:

    1)    $C_3 \subseteq \overline{C_2}$

    2)   $SC_1 \subseteq C_2$

∴   3)   $SC_1 \subseteq \overline{C_3}$

Due to (2) and (1), the following equivalence is valid:

$$(C_3 \subseteq \overline{C_2}) \leftrightarrow (C_2 \subseteq \overline{C_3})$$

Therefore, the syllogism considered may be reformulated as:

    1)    $C_2 \subseteq \overline{C_3}$

    2)   $SC_1 \subseteq C_2$

$$\therefore \quad 3) \quad SC_1 \subseteq \overline{C_3}$$

The conjunction of the two premises of the syllogism, once it is reformulated, is the following proposition:

$$(SC_1 \subseteq C_2) \wedge (C_2 \subseteq \overline{C_3})$$

Due to (4), the following proposition is valid:

$$\left((SC_1 \subseteq C_2) \wedge (C_2 \subseteq \overline{C_3})\right) \rightarrow \left(SC_1 \subseteq \overline{C_3}\right)$$

If both premises are true, then the proposition $(SC_1 \subseteq \overline{C_3})$ also is true. This proposition is the conclusion of the syllogism considered.

The above reasoning may be shown graphically as follows:

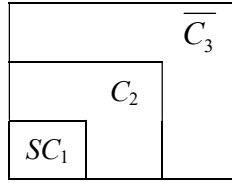

Figure 14. Inclusion diagram for the categorical syllogism in example 14.

Example 15

     1) No journalist is promiscuous.
     2) Some promiscuous persons are fugitives.
$\therefore$   3) Some fugitives are not journalists.

Consider the following sets:

$C_1$: fugitives;        $C_2$: promiscuous persons;       $C_3$: journalists
$SC_1$: some fugitives;   $SC_2$: some promiscuous persons

Given the notation introduced, the syllogism considered may be expressed as:

     1)   $C_3 \subseteq \overline{C_2}$
     2)  $SC_2 \subseteq C_1$
$\therefore$   3)  $SC_1 \subseteq \overline{C_3}$

Due to (2) and (1), the following equivalence is valid:

$$(C_3 \subseteq \overline{C_2}) \leftrightarrow (C_2 \subseteq \overline{C_3})$$

Due to (3), the following equivalence is valid:

$$(SC_2 \subseteq C_1) \leftrightarrow (SC_1 \subseteq C_2)$$

Therefore, the syllogism considered can be reformulated as:

      1)    $C_2 \subseteq \overline{C_3}$

      2)  $SC_1 \subseteq C_2$

∴   3)  $SC_1 \subseteq \overline{C_3}$

The conjunction of the two premises of the syllogism, once reformulated, is the following proposition:

$$(SC_1 \subseteq C_2) \wedge (C_2 \subseteq \overline{C_3})$$

Due to (4), the following proposition is valid:

$$\left((SC_1 \subseteq C_2) \wedge (C_2 \subseteq \overline{C_3})\right) \rightarrow \left(SC_1 \subseteq \overline{C_3}\right)$$

If both premises are true, then the proposition $(SC_1 \subseteq \overline{C_3})$ also is true. This proposition is the conclusion of the syllogism considered.

The above reasoning may be shown graphically as follows:

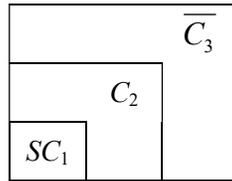

Figure 15. Inclusion diagram for the categorical syllogism in example 15.

Example 16

     1) No chess player is naïve.

     2) Some chess players are Russians.

∴   3) Some Russians are not naïve.

Consider the following sets:

  $C_1$: Russians;         $C_2$: chess players;        $C_3$: naïve persons

$SC_1$: some Russians;   $SC_2$: some chess players

Given the notation introduced, the syllogism considered can be expressed as:

$$1) \quad C_2 \subseteq \overline{C_3}$$
$$2) \quad SC_2 \subseteq C_1$$
$$\therefore \quad 3) \quad SC_1 \subseteq \overline{C_3}$$

Due to (3), the following equivalence is valid:

$$(SC_2 \subseteq C_1) \leftrightarrow (SC_1 \subseteq C_2)$$

Therefore, the above syllogism can be reformulated as:

$$1) \quad C_2 \subseteq \overline{C_3}$$
$$2) \quad SC_1 \subseteq C_2$$
$$\therefore \quad 3) \quad SC_1 \subseteq \overline{C_3}$$

The conjunction of the two premises, once the syllogism is reformulated, is the following proposition:

$$(SC_1 \subseteq C_2) \wedge (C_2 \subseteq \overline{C_3})$$

Due to (4), the following proposition is valid:

$$\left((SC_1 \subseteq C_2) \wedge (C_2 \subseteq \overline{C_3})\right) \rightarrow \left(SC_1 \subseteq \overline{C_3}\right)$$

If the two premises are true, then the proposition $(SC_1 \subseteq \overline{C_3})$ also is true. This proposition is the conclusion of the syllogism considered.

The above reasoning may be shown graphically as follows:

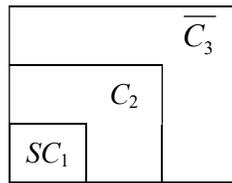

Figure 16. Inclusion diagram for the categorical syllogism in example 16.

Example 17

       1) All men are mortals.
       2) Socrates is a man.

$\therefore$    3) Socrates is a mortal.

This well-known syllogism is characterized by having, as its minor term, "Socrates", a being which cannot be thought of as a set. To apply the inclusion diagram approach to this syllogism, it is useful to consider the set $C_1$, to which Socrates belongs, with only one element: $C_1$ = (Socrates).

Consider the following sets:

$C_1$: (Socrates);      $C_2$: men;      $C_3$: mortals

Given the notation introduced, the syllogism considered may be expressed as:

    1)   $C_2 \subseteq C_3$

    2)   $C_1 \subseteq C_2$

$\therefore$   3)   $C_1 \subseteq C_3$

The second premise of the above syllogism is $(C_1 \subseteq C_2)$. Given that the set $C_1$ is included in the set $C_2$, any element belonging to $C_1$ also belongs to $C_2$. Consequently, the only element belonging to $C_1$ (Socrates) also belongs to $C_2$. Since $C_2$ is the set of mortals, Socrates, the element belonging to $C_2$, is mortal. It can be seen that $(C_1 \subseteq C_2)$, with the meaning specified for those sets, is a correct reformulation of the second premise of the syllogism considered, in its original form. For similar reasons, $(C_1 \subseteq C_3)$ is a correct reformulation of the conclusion of that syllogism, as it was in its original form.

The conjunction of the two premises is the following proposition:

$$(C_1 \subseteq C_2) \wedge (C_2 \subseteq C_3)$$

Due to (4), the following proposition is valid:

$$\big((C_1 \subseteq C_2) \wedge (C_2 \subseteq C_3)\big) \rightarrow \big(C_1 \subseteq C_3\big)$$

If both premises are true, then the proposition $(C_1 \subseteq C_3)$ also is true. This proposition is the conclusion of the syllogism considered.

The above reasoning may be shown graphically as follows:

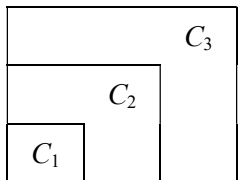

Figure 17. Inclusion diagram for the categorical syllogism in example 17.

Example 18

This example will serve to emphasize that determining the truth or falsehood of propositions about physical, biological or social facts, and the like, as are those studied by factual sciences, is beyond the scope and objective of logic. In contrast, logic does provide tools to determine whether reasoning of different types is correct or incorrect, or valid or not valid, as in syllogisms, for example.

     1) All cats are invertebrates.
     2) All invertebrates are gorillas.
∴   3) All cats are gorillas.

According to currently accepted zoological knowledge, both premises as well as the conclusion of the above syllogism are false. However, it will be shown that this syllogism is a valid form of reasoning, and therefore, formally correct.

Consider the following sets:

$C_1$: cats;     $C_2$: invertebrates;     $C_3$: gorillas

Given the notation introduced, the syllogism considered may be expressed as:

    1)   $C_1 \subseteq C_2$
    2)   $C_2 \subseteq C_3$
∴  3)   $C_1 \subseteq C_3$

The conjunction of the two premises is the following proposition:

$$(C_1 \subseteq C_2) \wedge (C_2 \subseteq C_3)$$

Due to (4), the following proposition is valid:

$$\big((C_1 \subseteq C_2) \wedge (C_2 \subseteq C_3)\big) \rightarrow \big(C_1 \subseteq C_3\big)$$

If both premises are true, then the proposition $(C_1 \subseteq C_3)$ also is true. This proposition is the conclusion of the syllogism considered.

This formally valid syllogism may be expressed graphically as follows:

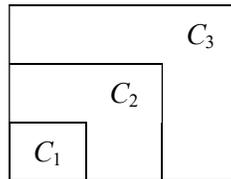

Figure 18. Inclusion diagram for the categorical syllogism in example 18.

## 5. Analysis of a categorical syllogism which is not valid

In this section, the following categorical syllogism will be analyzed:

1) Some professors are visionaries.
2) All poets are visionaries.

∴   3) Some poets are professors.

Consider the following sets:

$C_1$: poets;        $C_2$: visionaries;     $C_3$: professors
$SC_1$: some poets;                        $SC_3$: some professors

Given the notation introduced, the syllogism considered may be expressed as:

1)   $SC_3 \subseteq C_2$

2)    $C_1 \subseteq C_2$

∴   3)   $SC_1 \subseteq C_3$

The conjunction of the premises is the following proposition:

$$(SC_3 \subseteq C_2) \wedge (C_1 \subseteq C_2)$$

In this case, it is not possible to apply (4) to the conjunction of the premises. This is because in categorical syllogisms which are not valid, the conclusion is not a logical consequence of the premises. The fact that both premises are considered true does not imply that the conclusion must also be true.

If, for a given categorical syllogism, an inclusion diagram cannot be constructed, like those which have been presented for the 18 examples presented above of valid categorical syllogisms, then that categorical syllogism is not valid.

How can this type of inclusion diagram be characterized unambiguously as in the 18 examples given above? As follows: In all of these examples, one of the premises will represent a first set included in a second set, and the other premise will represent that second set included in a third set. The conclusion of the syllogism, in every case, is a proposition in which it is affirmed that the first set is included in the third set. Of course, any proposition that expresses the inclusion of a set in another set may be replaced, if appropriate, by an equivalent proposition, according to set theory. (To facilitate this type of operation, results (1), (2), (3), and (4) from set theory were presented above.)

## 6. Inclusion diagram for reasoning other than that of categorical syllogisms

For $n$ sets $C_1$, $C_2$, $C_3$, ..., $C_{n-1}$, $C_n$, such that between them there exist inclusion relations $(C_1 \subseteq C_2)$, $(C_2 \subseteq C_3)$, ..., $(C_{n-1} \subseteq C_n)$, the following obvious and intuitive generalization of (4) is valid: $\left( (C_1 \subseteq C_2) \wedge (C_2 \subseteq C_3) \wedge ... \wedge (C_{n-1} \subseteq C_n) \right) \rightarrow \left( C_1 \subseteq C_n \right)$ (5).

Of course, any change in the denominations of the sets $C_1$, $C_2$, $C_3$, ..., $C_{n-1}$, $C_n$, still preserves the validity of (5) because the mere change in the denomination of a set does not alter its nature, that is, the elements pertaining to that set.

Consider the following premises for reasoning [8]:

1) I greatly value everything that John gives me.
2) Nothing but this bone will satisfy my dog.
3) I take particular care of anything that I greatly value.
4) This bone was a present from John.
5) When I take particular care of anything, I do not give it to my dog.

It may be admitted that the presents from John are some of the things that John gives me. John may give me other things that are not presents from him. Thus, for example, the other things could be things that John lends me or sells me. It is reasonable, therefore, to add this sixth premise to the first five.

6) The presents from John are things that John gives me.

What can be concluded from the above premises? That conclusion will be provided below.

Consider the following sets:

$C_1$: set of objects that I value highly.
$C_2$: set of objects that John gave me.
$C_3$: set with only one object: the bone.
$C_4$: set of objects that satisfy my dog.
$C_5$: set of objects that I take particular care of, or worry about.
$C_6$: set of objects that are gifts from John.
$C_7$: set of objects that I will not give my dog.

Using the above notation, we can reformulate the premises of the reasoning considered as follows:

1)  $C_2 \subseteq C_1$
2)  $C_3 = C_4$
3)  $C_1 \subseteq C_5$
4)  $C_3 \subseteq C_6$
5)  $C_5 \subseteq C_7$
6)  $C_6 \subseteq C_2$

The conjunction of the premises 4), 6), 1), 3) and 5) is the following proposition:

$$(C_3 \subseteq C_6) \wedge (C_6 \subseteq C_2) \wedge (C_2 \subseteq C_1) \wedge (C_1 \subseteq C_5) \wedge (C_5 \subseteq C_7)$$

If (5) is applied to the preceding proposition, the following proposition is obtained:

$$\big((C_3 \subseteq C_6) \wedge (C_6 \subseteq C_2) \wedge (C_2 \subseteq C_1) \wedge (C_1 \subseteq C_5) \wedge (C_5 \subseteq C_7)\big) \rightarrow (C_3 \subseteq C_7)$$

If the premises 4), 6), 1), 3) and 5) are true, then the proposition $(C_3 \subseteq C_7)$ also is true. Given the meanings attributed to the sets $C_3$ and $C_7$, $(C_3 \subseteq C_7)$ can be expressed in common language as: I will not give this bone to my dog. If it is accepted that premise 2) is true, then in the proposition $(C_3 \subseteq C_7)$, $C_4$ may be used in place of $C_3$. Thus the following proposition is obtained: $(C_4 \subseteq C_7)$. In common language, this proposition, which is the conclusion of the reasoning considered, can be expressed as follows: My dog will not be satisfied.

The reasoning considered has been shown graphically in figure 20 using an inclusion diagram.

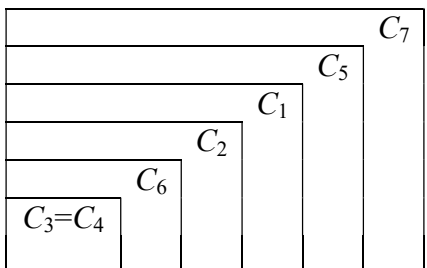

Figure 20: Inclusion diagram for the reasoning considered in this section.

## 7. Discussion and prospects

The critical discussion of the validity of a categorical syllogism can be carried out using set theory, as done above, or using predicate calculus, for example. For this purpose, graphic representations, such as the well-known Venn diagrams or the inclusion diagrams introduced here, are unnecessary. However, these graphic representations can contribute to a clear understanding and intuitive comprehension of whether categorical syllogisms are correct.

The use of inclusion diagrams for the analysis of the validity of categorical syllogisms can serve as an alternative to Venn diagrams for this purpose; or at the very least, this type of inclusion diagrams provide another perspective from which to approach the issue.

For cases in which not three, but four sets are involved, Venn diagrams must be modified, using areas bounded by ellipses and not by circumferences, to represent the different sets. In addition, it is not too feasible to use Venn diagrams for reasoning requiring more than four sets [9].

In contrast, inclusion diagrams are easily applicable in different types of arguments involving more than four sets. In section 6, attention was given to reasoning involving six sets.

The scope and limitations of inclusion diagrams, as well as some of their variants, will be further discussed elsewhere, as tools for the analysis of diverse types of reasoning different from categorical syllogisms.

## References

[1] Quine, W.V. (1982). *Methods of Logic*. 4th ed. Cambridge: Harvard University Press, pp. 98-101, 102-108, 109-113.

[2] Moktefi, A. and S.-J. Shin, eds. (2013). *Visual Reasoning with Diagrams*. New York: Birkhäuser, Springer.

[3] Allwein, G. and J. Barwise, eds. (1996). *Logical Reasoning with Diagrams*. Oxford: Oxford University Press.

[4] Anderson, M., B. Meyer, and P. Olivier, eds. (2002). *Diagrammatic Representation and Reasoning*. London: Springer.

[5] Van Dyke, F. (1995). "A visual approach to deductive reasoning," *The Mathematics Teacher*, 88, 6: 481-486, 492-494.

[6] Nakatsu, R. T. (2010). *Diagrammatic Reasoning in AI*. Hoboken, NJ: Wiley.

[7] Sasakura, M. (2001). "A visualization method for knowledge represented by general logic programs," *Proceedings Fifth International Conference on Information Visualization (IEEE)*.

[8] Carroll, L. (1896). *Symbolic Logic*. 2nd ed. London: Macmillan, p. 93.

[9] Quine, W.V. (1982) *Methods of Logic*. 4th ed. Cambridge: Harvard University Press, p. 110.